\documentclass[times,Royal]{sagej}

\usepackage{moreverb,url}

\usepackage[default]{lato}
\usepackage[T1]{fontenc}

\usepackage[colorlinks,bookmarksopen,bookmarksnumbered,citecolor=red,urlcolor=red]{hyperref}
\newcommand\BibTeX{{\rmfamily B\kern-.05em \textsc{i\kern-.025em b}\kern-.08em
T\kern-.1667em\lower.7ex\hbox{E}\kern-.125emX}}

\begin{document}

\runninghead{Heroy et al.}

\title{Are neighbourhood amenities associated with more walking and less driving? Yes, but only for the wealthy.}
\author{Samuel Heroy\affilnum{1,2}, Isabella Loaiza\affilnum{3}, Alex Pentland\affilnum{3} and Neave O'Clery\affilnum{1,2}}

\affiliation{\affilnum{1}Bartlett Centre for Advanced Spatial Analysis, University College London, UK\\
\affilnum{2}Mathematical Institute, University of Oxford, UK\\
\affilnum{3}MIT Media Lab, Massachusetts Institute of Technology, MA, USA}

\corrauth{Samuel Heroy, Bartlett Centre for Advanced Spatial Analysis, University College London, London W9 1TJ, UK.}

\email{s.heroy@ucl.ac.uk}

\begin{abstract}
Cities are home to a vast array of amenities, from local barbers to science museums and shopping malls. But these are inequality distributed across urban space. Using Google Places data combined with trip-based mobility data for Bogot\'a, Colombia, we shed light on the impact of neighbourhood amenities on urban mobility patterns. Deriving a new accessibility metric that explicitly takes into account spatial range, we find that a higher density of local amenities is associated a higher likelihood of walking as well as shorter bus and car trips. Digging deeper, we use a sample stratification framework to show that socioeconomic status (SES) modulates these effects. Amenities within about a 1km radius are strongly associated with a higher propensity to walk and lower driving time only for only the wealthiest group. In contrast, a higher density of amenities is associated with shorter bus trips for low and middle SES residents. As cities globally aim to boost public transport and green travel, these findings enable us to better understand how commercial structure shapes urban mobility in highly income-segregated settings. 
\end{abstract}





\maketitle

\section{Introduction}

Playing a key role in the urban economy, urban business amenities (``street commerce") are thought to attract people to live and spend time in vibrant cities. Glaeser and coauthors pioneered the idea of the 'consumer city' - suggesting that American cities' growth in modern times is largely due to consumption-related benefits of living in a city rather than wage premiums \cite{glaeser2001consumer}. Their seminal paper made the case that high amenity cities have grown faster than low amenity cities, and that urban rents have grown faster than urban wages \cite{glaeser2001consumer}. Recent work in a variety of global settings has expanded on this finding - showing generally that cities with more amenities and more amenity diversity tend to be more attractive to tourists \cite{carlino2019beautiful}; and neighbourhoods with more amenities draw more visits and also have higher real estate values \cite{chong2020economic, oner2017retail}. However, this evidence has recently been subject to some pushback in the Global South, with recent evidence from Colombia suggesting that amenities have not had a significant effect on urban growth patterns \cite{duranton2016agglomeration}, signaling further need for research on amenities in developing contexts.

In particular, ``walkability" - the ease with which residents can walk to nearby stores, parks, schools, shops, cafes, and other amenities - has emerged as an important and popular feature of cities and individual neighborhoods \cite{quastel2012sustainability, xia2018assessing, rauterkus2011residential}. Most recently, the COVID-19 pandemic has renewed and placed further focus on neighborhood amenities, and in particular various ideals of a post-pandemic walkable neighbourhood \cite{15minute1, 15minute2, moreno2021introducing, guzman2021covid}. 

Urban planning with walkability in mind has emerged as a policy priority, and even entered the public discourse in countries of varying income levels. For instance, Anne Hidalgo's successful mayoral campaign in Paris popularized the ``15 minute city" concept in which each arrondissement would have an accessible array of amenities \cite{moreno2021introducing, 15minute1}. This concept has subsequently gained popularity worldwide including in Latin America. More broadly, cities around the world are moving more towards mixed-use development - which brings residential housing closer to commercial amenities and away from purely residential development - so as to increase walking and reducing car travel. For instance, the World Bank regularly advocates for mixed use planning in its master planning guidance\cite{wb_mp}, while the Inter-American Development Bank also frequently cites it as a development priority \cite{idb}. The UK ministry of Department for Levelling Up, Housing \& Communities identifies mixed use planning as ``the way forward" and a key priority for its 10 billion pound development plan \cite{bbbhs}. Latin American cities are similarly embracing mixed use \cite{molina2014expansion} while U.S. cities are slowly moving away from residential zoning \cite{montgomery}. 

There is particular interest in creating walkable communities as a key component of urban revitalization plans with a social sustainability focus. For example, the \$170 million (USD) `Parques del R\'io' project in Medell\'in is a megaproject aiming to inspire mixed use development by revitalising select areas, highway relocation, infrastructure development and improved riverside connectivity. Planners envision that the project will lead to a revaluation of riverside single use land, which will also be targeted for mixed use investments in housing and commerce \cite{colombiaparques}. In particular, there is focus on increasing amenity accessibility for the cities' blue collar population, and developing low income areas that formerly were associated with violence \cite{colombiaparques2}. 

At the same time, there are challenges to creating walkable mixed use communities in low and high income areas alike \cite{grant2002mixed}. Generally, mixed use projects face obstacles due to both cultural and economic reasons. Mixed use investments are expensive, and frequently require tax incentives to be viable. Mixed use planning in low-income areas also faces difficulties on account of more limited investment \cite{freemark2018challenges}. Second, mixed use developments frequently face resistance from local communities, which may be either reluctant to see neighbourhood change and/or are concerned about increasing real estate prices (e.g., gentrification). For instance, critics of Medell/'in's development projects cite both their heavy price tag, and their disruption to the `location, land, and social capital' of low income communities \cite{anguelovski2019grabbed} as well as the security of upper income communities \cite{guardianparques}.

Yet, behind all of this debate and hype surrounding walkable communities, there lies a key question: do amenities encourage more walking and less driving in reality? Many scholars emphatically claim that having nearby amenities - like compact development in general \cite{kramer2013our, ewing2001travel, ewing2015varying, ewing2017does} - encourages people to drive less and walk more \cite{ellder2020kind, ellder2020local, haugen2013divergent}. Additionally, there are a variety of other important factors that have been cited as decreasing car travel and increasing walking, for instance transport structure and pedestrian safety \cite{koschinsky2017walkable, ewing2001travel}. Additionally, there are important contextual factors that influence the impact of amenities on transport behaviour. For instance, evidence from Sweden indicates that the effect of amenities on trip behaviour is greater in rural villages vs. city centres \cite{ellder2020local}. What is less clear in the literature, however, is the impact of income or socio-economic status on the relationship between amenities and mobility. In other words, does the presence of local amenities influence travel decisions more for poorer or richer residents? 

Socioeconomic status matters for transport decisions, and so it stands to reason that amenities may have different effects on transport decisions for different income groups. This issue is of key importance for development projects which aim to increase active transport and reduce car travel. For instance, increasing the amenity supply via commercial development in a wealthy suburb might have different effects on local transport decisions as compared to a blue collar neighbourhood or an informal settlement. 

Here, we build on existing literature \cite{ellder2020kind, ellder2020local, haugen2013divergent} to shed light on the relationship between amenities and mobility behaviour, with a specific focus on how socioeconomic status modulates this relationship. In particular, we use amenity location data to construct a neighborhood-level measure for accessibility to amenities. Importantly, this measure is adaptable to various spatial scales so as to capture for instance walking vs. driving accessibility to amenities. Using survey data on non-work trips, we investigate the overall effect of amenity accessibility - alongside various controls - on residents' travel duration across various travel modes, and as well on their tendency to make walking trips. Then, using a simple sample stratification framework, we investigate how these effects differ for varying socioeconomic groups. We also vary the spatial scale of the accessibility measure in order to better understand how distance to amenities influences our results. 

Most related to our work, the majority of existing research on the relationship between amenities in transport behaviour is to our knowledge concentrated in Sweden, while most analysis of the relationship between the built environment and transport decisions generally focuses on the USA, Europe, or China, with relatively few papers focused on developing settings. Due to rapid rise in automobile ownership \cite{yanez2019urban}, there is a pressing need to study the determinants of transport behaviour in Latin America in order to understand the extent to which built environment interventions might have the capacity to mitigate rising car use.

We chose Bogot\'a as our study region on account of its rich variation in both urban form and mobility. As in many other Latin American metropolises, Bogot\'a has high socioeconomic spatial inequality \cite{morales2008quality} and is characterized by a large number of densely populated low \emph{estrato} (socioeconomic stratum) areas. Car ownership varies across the city, with lower income areas having approximately 1 car for every 5 households while higher income areas have 1.3 cars for every household \cite{guzman2017urban}. One of the largest cities in the world with no intra-urban rail system, traffic in Bogot\'a is regularly cited as among the worst \cite{tomtom}, even though it has improved considerably since the introduction of bus rapid transit \cite{hidalgo2013transmilenio} as well as congestion pricing (the well-known "Pico y placa'' programme). Like density, congestion and high travel time disproportionately affect low stratum residents \cite{guzman2017urban, origindest}. Much study on mobility inequality especially in Bogot\'a is concerned with spatial separation between residents and jobs \cite{guzman2017urban, guzman2017assessing, guzman2018accessibility, bocarejo2016accessibility}. These studies point towards the fact that the city is monocentric with regards to the concentration of employment, and SES decays linearly with distance to the city center, giving rise to burdensome home-work separation for poorer citizens. While this is understandably a key area of concern, work-related travel constitutes less than 1/3 of all trips in metropolitan Bogot\'a \cite{origindest}, signaling a need for further study of non-work travel behavior. 
 
\section{Literature review}

Since at least the 1980s, there has been debate about whether compact, dense development is good for cities, but most planners today view it to be overall beneficial for urban mobility \cite{ewing1997angeles, ewing2001travel, ewing2017does, ewing2015varying}. Compact urban development, as opposed to sprawl, is thought to reduce residents' overall car travel time and distance travelled as well as to increase their tendency to make walking or biking trips ('active transport'). How is compact development measured? Usually, studies on the relationship between the built environment and travel behaviour are focused on the impacts of the five\footnote{More have been proposed but these ones are well established.} `Ds of the built environment' which include:

\begin{itemize}
    \item diversity in land use;
    \item density, either in terms of population or jobs;
    \item destination accessibility, e.g. distance to the city centre;
    \item design of the street network, often measured in terms of the density of street intersections; and
    \item distance to transit.
\end{itemize}

The effect size of compact development on car travel has come under much debate - articles on this topic include some of the most read and cited articles ever in planning journals \cite{ewing2001travel}. In particular, Mark Stevens \cite{stevens2017does} argues via meta-regression (a form of cross-context analysis) that the (negative) effect size of compact development on driving distance is actually relatively modest, specifically finding what he considers to be a relatively low elasticity (-0.22) of population density on driving distance (and even lower elasticities for other factors except distance to the city centre). Susan Handy, Reid Ewing, Robert Cervero and others \cite{handy2017thoughts, ewing2017does} argue to the contrary that compact development is very important, both proposing alternative results and also disputing Stevens' interpretation of his low elasticities. In perhaps the most well-known effort to establish causation, Susan Handy and coauthors showed that changes in the built environment over time do lead to changes in transport behaviour \cite{handy2005correlation}.

Recently, there has been a trend towards more localised analysis of neighbourhood-level factors that influence mobility behaviours. Traditional measures of the built environment are easy to conceptualize and doubtlessly have practical planning utility, but these are hard to measure at high precision in a localised intra-urban setting, highlighting a need for more spatially resolved data \cite{handy2002built}. In particular, recent evidence points towards an important role for local amenities in residents' travel patterns \cite{haugen2013divergent, ellder2020kind, ellder2020local}. We now discuss in more detail the results of three studies that leverage Swedish micro-data on amenity location in combination with travel surveys. 

First, consistent with the literature on compact development, Ellder and coauthors \cite{ellder2020kind} show that having more amenities nearby is associated with reduced travel time and more active transport, even when controlling for the five D's of the built environment. At the local scale, amenities have strong impact on vehicle kilometers traveled as well as the tendency to use active transport, while variables such as street network design, density, and diversity are less important. 

Second, Haugen and coauthors \cite{haugen2013divergent} point to the importance of spatial scale. While higher `local accessibility' (amenities within 1-5 km) is associated with less distance traveled, higher `regional accessibility' (more amenities within 50 km) is associated with greater distance traveled. This study highlights the need for a better understanding of spatial scale in accessibility - does distance to amenities matter? 

Third, Ellder and coauthors \cite{ellder2020local} show that the effect of amenities on travel behaviour varies according to the setting. In a wider, regional analysis, the authors find that amenities have distinct effects on travel behaviour in urban vs. rural settings. Distance to a few nearby amenities (especially essential ones like grocery stores) plays a more important role in driving time in a rural setting vs. a city setting wherein there are more choices. 

However, urban vs. rural is of course not the only type of context that is likely to affect people's dependence on amenities. For instance, there is vast literature investigating travel behaviour by gender, educational attainment, age and socioeconomic status \cite{paul2015walking, duchene2011gender}. There are reasons why varied demographic groups might be affected by amenities differently with regards to their transport behaviour. For instance, older and/or less physically active people may be less willing to walk 1 km to an amenity than their peers.

Here we focus on the role of SES in the relationship between amenities and mobility. Why might accessibility to amenities affect travel decisions differently based on SES? Or, equivalently, why would the effect of amenities on mobility behaviour not be generalisable across SES groups? People make urban travel decisions based primarily on time and money. We might naively expect that poor people aim to lower travel costs based on monetary considerations, while the rich are more inclined to minimise trip duration \cite{boarnet2001travel, mondschein2018persistent} or be motivated by other factors like pedestrian safety. Hence, we might expect that wealthy people's decision to walk or drive is highly motivated by proximity to amenities - e.g., they are more willing to walk when amenities are closeby and walking is more convenient than driving. Moreover, we might expect that lower income people are less sensitive to the proximity of amenities - and are willing to walk longer distances to amenities rather than spend money on fuel or bus fare.

There is much existing literature regarding the effect of SES on travel mode choice. Ewing and Cervero \cite{ewing2001travel} surmise that - unlike in the case of travel distance - SES is perhaps more important than built environment factors when it comes to travel mode choice. The relationship between SES and walking can vary according to location and the type of trip. For instance, in US cities the rich walk more than the middle class but less than the poor \cite{mondschein2018persistent}, possibly due to recreational walking \cite{agrawal2007extent}. This relationship has also been affected by the pandemic, as wealthier groups have increased their leisure walking relative to pre-pandemic levels \cite{hunter2021effect}. In Colombian cites, it is clear that lower income people make far more trips by walking \cite{guzman2017urban} and as well make more local trips \cite{marquet2017local}. The latter finding has been used to argue that lower income people have a closer relationship to their immediate environment \cite{marquet2017local}. Overall, lower SES is certainly associated with more walking, but it is not clear how this relationship depends on amenities.

There is also some previous work aimed at understanding how SES and urban structure jointly affect travel distance. Ewing and Cervero \cite{ewing2001travel} empirically find that factors related to the built environment are more important for residents' travel distance than SES. More recently, the availability of mobile phone data has further illuminated the relationship between SES, urban structure and mobility. In a recent cross-city analysis focusing on US and Brazilian cities, Barbosa \emph{et al} \cite{barbosa2021uncovering} found that when the cities have low public transport availability and the wealthy live further from amenities, the rich travel further than the poor. When cities have more public transport and services are more distributed, this discrepancy in mobility disappears. This finding supports the hypothesis that wealthy residents depend on amenities with regards to their mobility, but does not necessarily suggest that other income brackets depend less so or differently on amenities. We argue that a neighbourhood-level analysis can help to fill this gap. 

Here, we utilise a sample stratification framework to investigate the effects of amenity accessibility on mobility behaviour for low, middle, and high SES groups. This type of framework has been used extensively, for example, to study the impact of the local environment on health. Conducting a literature review of these studies (focused on ``socially differentiated vulnerability to place effects"), Vall\'ee and colleagues \cite{vallee2021everyday} find that in many contexts spatial accessibility to services has a strong effect on health outcomes for poorer and lower education people, yet little or no effect on many types of health outcomes for wealthy people. In the setting of mobility behaviour, our expectation is the opposite - that wealthier people are more dependent on their local neighbourhood. 
 
\section{Data}
\label{sec:data}
\subsection{Study area}
Our study focuses on the metropolitan area of Bogot\'a, including the 19 urban localities of the municipality of Bogot\'a itself as well as five contiguous municipalities in the metropolitan area (Soacha, Mosquera, Ch\'ia, Funza and Cota). In other statistical delineations of the metropolitan area \cite{guzman2017urban,guzman2017city}, 12-17 municipalities are included. However, many of these municipalities are small and/or are not fully integrated into Bogot\'a \cite{guzman2017city} and hence we exclude them. We perform our analyses at the level of urban zonal planning units (Unidades Territoriales de An\'alisis de Movilidad), which we henceforth refer to as zones. There are 954 of these in the study area, and they have have a median area of 0.38 km$^2$. The next level of aggregation is cells - there are 121 cells in the study area with median area 3.58 km$^2$.

\subsection{Amenities}
Our main source data on amenities come from the Google Places API. Downloaded in March-April 2021, the data contains the location, name, and class for a wide range of amenities in the study area. Google Places data is sourced from a combination of publicly available data, crowd sourcing, and licensed data from third parties\footnote{https://support.google.com/business/answer/2721884?hl=en}. Amenities from Google Places are most aptly characterized as focusing on customer-facing businesses and neighborhood points of interest. These fall within nearly 100 categories, but we focus specifically on 14 amenity classes which we choose in order to include a breadth of different activities including retail, health, food/beverage consumption, physical exercise, haircare, personal finance, and libraries (see Table~\ref{tab:my_label}). We extract 59,277 amenities in total across these types in the entire study area.

\begin{table}[]
    \centering
    
    \begin{tabular}{c|c}
         Amenity class $c$ & Total count ($\sum_z A_z^c$)\\
         \toprule
         Physiotherapists&147  \\
         Libraries&198\\
         Bicycle stores & 839\\
         Veterinarians & 1884\\
         Gyms & 2444\\
         Electronics stores & 3304\\
         Banks & 3699\\
         Dentists &3985\\
         Beauty salons & 5035\\
         Pharmacies & 6461\\
         Cafes &6995\\
         Home goods stores &7782 \\
         Clothing stores &8170\\
         Convenience stores &8334 \\
    \end{tabular}
    \caption{Number of amenities across the study area}
    \label{tab:my_label}
\end{table}

As noted by \cite{hidalgo2020amenity}, there are of course deficiencies in this dataset. Firstly, some amenities may be multiply counted due to multiple labels (e.g. convenience stores that are also electronics stores), and some places may have labels that are incorrect (e.g. a park field labeled as a gym). Additionally, the dataset is not dynamic and some places are mislabeled as operational when they are closed (especially given widespread closures during the COVID-19 pandemic). Nevertheless, the dataset is widely thought to have more extensive coverage and higher accuracy than related datasets (e.g. OpenStreetMaps, Foursquare) \cite{safegraph,comparison}.

In a setting where many businesses do not have business licenses and so are not registered with official directory of establishments \cite{straulino2021uncovering}, the Google Places dataset has clear advantages over the official registry (i.e. the directory of establishments held by the local chamber of commerce). For example, just 1165 firms are registered as retail firms not at a residential address (sector: ``Retail trade in non-specialized stores" (CIIU 471)), yet there are 8,240 convenience stores in the Places database within the Bogot\'a municipality. For this reasons, we consider the Places dataset to be more representative than official data at the local level of commercial business activity (``street commerce"). 

\subsection{Socioeconomic stratum}
In order to proxy for socio-economic status, we use granular data on socio-economic `strata' \cite{DANE} constructed by the national government via surveys of urban residential dwelling conditions. This data is provided in the form of survey indices for each census block, with values ranging from 1 to 6: 1 denotes very poor quality housing (often informal), a majority of residents live in mid-stratum areas 2-4, while 5 and 6 include the most expensive areas of the city. Stratum tends to be highly spatially autocorrelated, owing to socio-economic segregation \cite{morales2008quality}. Strata are assigned to census blocks, but we use geographically weighted spatial averaging to estimate average strata at the zone level.

\subsection{Mobility survey}
In order to study the effects of amenities on residents' trip behaviour, we utilise the 2019 Encuesta de Movilidad survey which is a resident/household-level survey of 21,208 respondent households throughout the metropolitan area. The dataset covers a total of 134,498 trips and includes information on trip purpose, modality, origin/destination zone, time of day, time of week, and time elapsed. 

We use the survey data in two instances. First, we use the trip survey to estimate the spatial distribution of homes and jobs as in \cite{guzman2018accessibility}. With regards to jobs, this has an advantage over official employment statistics in that the survey data encompasses both formal and informal jobs. Because the sampling and weighting procedures may not accurately quantify the number of homes or jobs precisely at the zonal level, we aggregate these numbers to the cell level. At this level, we geolocate over 9.5 million homes and 2.5 million jobs across the study area. 

Secondly, we use the survey data to analyze effects on travel behavior for amenity-related trips originating from the respondents' home zones, focusing on trip duration and trip modality. Trip modality can broadly be distinguished as walking, cycling, driving (car/motorcycle), bus (including many different bus categories), taxi, informal transport, and a few other categories. The survey includes information related to trip purpose and mode. These purposes include work, work-related business, school/university, visiting friends/family, returning home, looking for someone, looking for a job, helping someone - and what we consider as amenity-related travel: receiving health attention, looking for something, going out to eat/drink, shopping, errands, recreation/culture, religious activities, and exercise/sports. Finally, because the survey is matched to respondent/household level characteristics, we can control for individual-level demographic factors, including socioeconomic stratum, age, gender, and whether the household has a motorized vehicle. 

\begin{figure}
    \includegraphics[width=\textwidth]{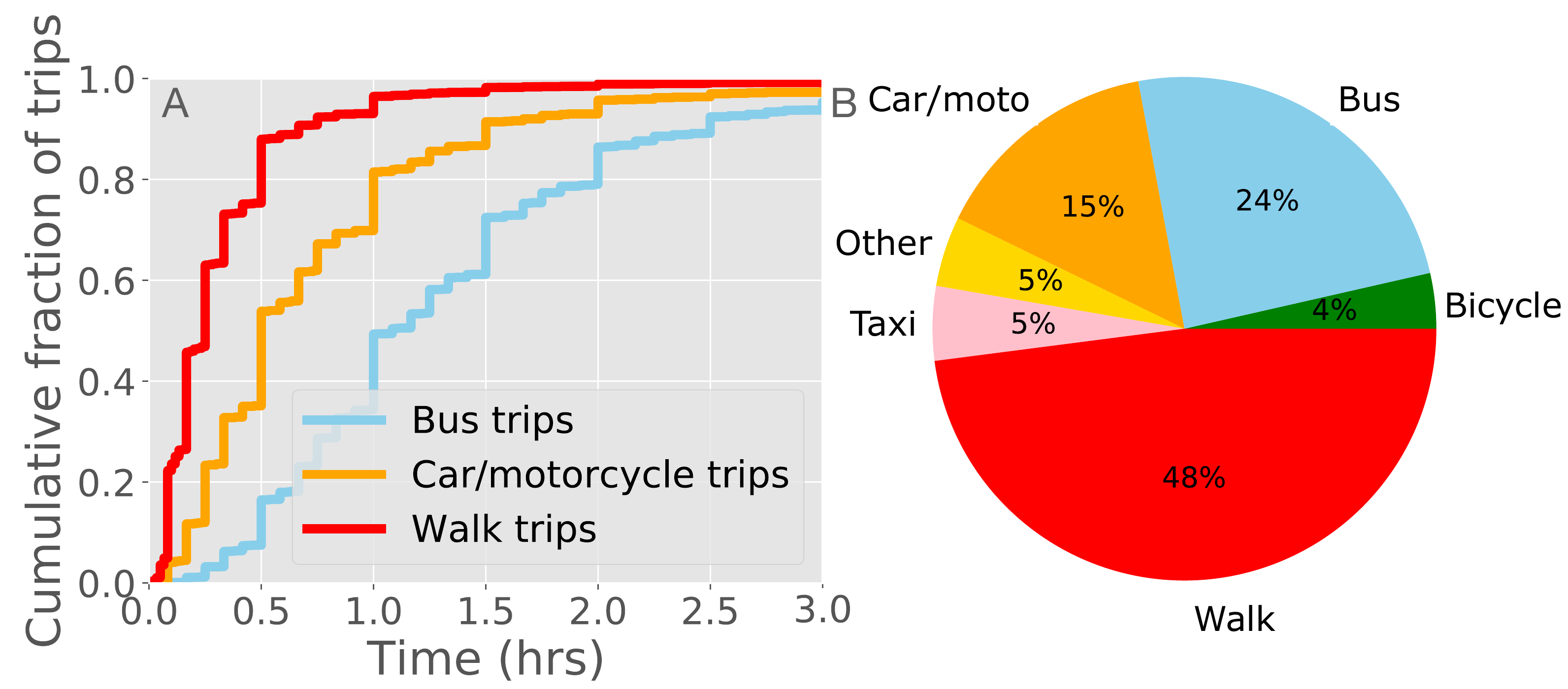}
    \centering
    \caption{{\bf Summary of dependent trip behaviour variables.} Subfigure A displays the distribution of time per trip across the three types of modes we study. For instance, about half of car trips take less than or equal to half an hour, while only about $20\%$ of bus trips are this short. Subfigure B displays the distribution of trip modes for the particular non-work trip purposes we study.}
    \label{figdv}
\end{figure}
 
\section{Variables and methods}

We begin by computing at the zone-level the count of amenities for each class. We denote this $A_z^c$, where $z$ denotes the zone and $c$ is the amenity class. 

\begin{figure}
    \centering
    \includegraphics[width=.8\textwidth]{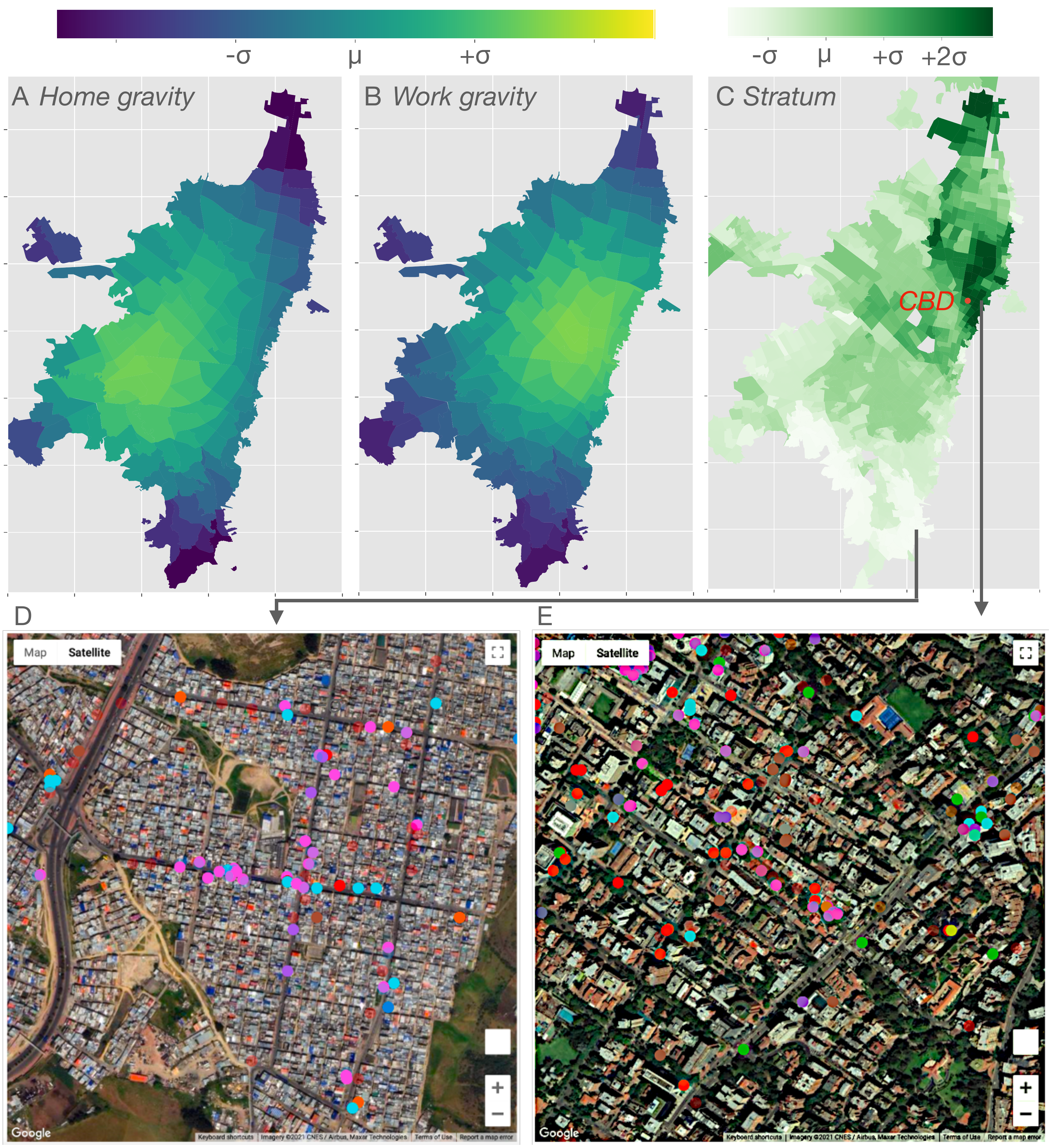}
    \caption{{\bf Overview of the city's urban form and SES structure} Subfigures A and B show the (gravity model) intensity of homes and jobs respectively. Subfigure C shows the stratum of each neighbourhood, a proxy for socioeconomic status, in addition to the city's generally recognized city centre. Subfigures D and E contrast the building density and amenity intensity of low and high stratum areas using satellite imagery. Lower stratum areas are home to many convenience stores (pink points), but lack the diversity of upper stratum areas. }
    \label{fig_dists}
\end{figure}

We are interested in the relationship between the spatial distribution of each amenity class and the distribution of homes, jobs and other socio-economic phenomena. We estimate the following equation for each amenity class $c$:  
\begin{small}
\begin{equation}
log(A_z^c)= 
\beta_0^c+\beta_h^c homes(grav)_z+\beta_w^c jobs(grav)_z+\beta_s^c stratum_z+\beta_d^c proxCBD_z.
\label{eq:poisson}
\end{equation}
\end{small} 
The first two terms are gravity-based intensity measures to quantify the concentration of homes and jobs at the cell level. Specifically, we deploy the exponential gravity formulation rather than raw counts in order to account for spatial dependencies between nearby zones as in \cite{sevtsuk2014location}. We display the spatial distribution of these densities in Figure \ref{fig_dists} A and B.

The latter two terms are chosen to reflect the tendency of certain amenities to concentrate in wealthier areas and to be centrally located in the city. In particular, we deploy a proxy for socioeconomic stratum, denoted $stratum$, which corresponds to government-delineated assessments of local living conditions \cite{DANE}. We also compute the (negative) intercentroidal distance in $km$ from each zone to the city's central business district (Chico Lago), denoted $proxCBD$ (we note that Bogot\'a is frequently labeled as a monocentric city \cite{guzman2017urban}). We show the spatial distribution of stratum and the location of the CBD in Figure \ref{fig_dists} C.
\subsection{Amenity accessibility metric}

In order to capture the proximity and density of the local amenity supply, for each class we compute the gravity accessibility of amenities at the zone level:
\begin{equation}
    access^c_z(\lambda)\doteq\sum_{z'=1}^N A_{z'}^c e^{- d_{zz'}/\lambda},
    \label{eqn:ind_access}
\end{equation}
which is frequently used to map out spatial distributions of accessibility for individual amenities (see review of amenity accessibility applications in \cite{nyerges2011sage}). 
The parameter $\lambda$ here sets the distance decay in importance of nearby amenities - a low $\lambda$ indicates that only nearby amenities contribute effectively, while higher $\lambda$ increases the relative contribution of further away amenities (see Fig.~\ref{fig_scales}). Here, we do not immediately choose a specific distance-decay parameter $\lambda$ a priori, because the scale of preference for localised amenities will vary with the application (e.g. walking vs. driving trips). 

While Eqn.~\ref{eqn:ind_access} can be used to capture the accessibility of amenities of a particular class, we aim to develop a composite measure which captures the accessibility across the range of classes. In order to do this, we simply normalise the individual amenity accessibilities, and then take the mean across amenity classes in order to obtain the mean share of accessibility to amenities:
\begin{equation}
     access_z(\lambda)\doteq\frac{1}{C}\sum_{c=1}^C \frac{access_z^c (\lambda)}{\sum_{z'=1}^N access_{z'}^c (\lambda)}.
     \label{eqn:ind_access_sum}
\end{equation}
This two-step approach has the advantage of capturing accessibilities to multiple amenity classes. For instance, not differentiating by class as we do in step one might give way to a high accessibility value in zones wherein one type of amenity is very well-represented, but others are sparse. An alternative way to capture accessibilities to various amenities would be to use an entropy or diversity measure. However, our proposed metric has the advantage of also utilising a gravity-based approach to proportionally weight nearby amenities, enabling us to compute the accessibility of a zone as a function of spatial scale, whereas diversity metrics (that we know of) typically account for amenities just within a particular spatial area. Our metric is quite similar to a select number of similar recently developed metrics that capture composite or integrated accessibility to various urban amenities \cite{ashik2020towards,li2021accessibility}, but our use of it differs in that we allow the distance decay parameter to vary in order to take into account various spatial ranges.

\subsection{Urban mobility and amenity accessibility}

In our main analysis, we examine the effect of amenity accessibility on urban mobility, focusing on trips that relate to use or patronage of urban amenities as described above. 

Specifically, we investigate the effect of amenity accessibility on two important dimensions of respondents' travel behavior:
\begin{enumerate}
    \item Trip modality, specifically the likelihood to make a trip by one particular mode as opposed to others, denoted by $Pr(mode=m)$.
    \item Trip duration in minutes, denoted $duration$\footnote{Exactly four trips have a reported time of $0$ minutes - to these we add one minute.}
\end{enumerate}
Easy to extract from trip survey data, these variables are traditionally used to capture everyday mobility patterns, and are similar to those considered by \cite{ellder2020kind, stevens2017does, ewing2001travel, ewing1997angeles, schwanen2002microlevel, cooke1999sample} and other studies. Another common metric deployed in the literature is trip distance. While both distance/duration are subject to reporting error, we use travel duration as in e.g., \cite{ewing2018does} instead of distance as the latter is not reported at a sufficiently granular level in our dataset. Additionally, travel duration is especially pertinent in a highly congested intra-urban setting in which distance might not well proxy for travel time. A general summary of our trip variables may be found in Fig.~\ref{figdv}. 

We deploy two related sets of models, one for the probability of mode choice (selection) and the other for trip duration (regression) according to various trip modes. It has been observed that performing selection/regression independently introduces bias in the regression model due to selectivity, and therefore we use Heckman's correction \cite{heckman1979sample} in order to fit both models jointly, in similar fashion to \cite{cooke1999sample, schwanen2002microlevel}.

Specifically, for each trip mode, we model mode selection using a probit model \footnote{Technically, it would be preferable to formulate this as one multinomial selection/regression 2-step problem, rather than separate binary selection/regression models. However, this modelling framework is more difficult to implement and may suffer from fitting difficulties. Secondly, the primary difference between individual binary selection models vs. one multinomial model is that in the latter case, the individual predicted probabilities for each mode choice sum up to one. As we are interested primarily in comparing regressor coefficients in the selection stage rather than prediction, this difference is not likely to meaningfully affect the results. For these reasons, we use separate binary selection/regression models.}:
\begin{footnotesize}
\begin{equation}
\label{model1}
Pr(mode_t=m)=\Phi\left(\alpha+ \beta \log \; access(\lambda)_{z(t)} +  (spatial\;controls)_z+ (trip\;controls)_t\right),
\end{equation}
\end{footnotesize}
where $t$ corresponds to a trip, where $z(t)$ refers to the zone in which the trip takes place, and $\Phi$ represents the standard normal distribution function.

These estimates for the probability of choosing mode $m$ are then transformed via the inverse mills ratio into a variable $\Lambda_t$, which is implicitly a function of the independent variables in Eq.~\ref{model2}. Simultaneously with fitting Eq.~\ref{model2}, we model trip $duration$ for trips of mode $m$ (``uncensored'' trips) according to:
\begin{footnotesize}
\begin{equation}
\label{model2}
    \log\; duration_t=\alpha+ \beta \log \; access(\lambda)_{z(t)} + (spatial\;controls)_z+ (trip\;controls)_t+ \beta_\Lambda \Lambda_t.
\end{equation}
\end{footnotesize}
The coefficient on the inverse mills ratio necessarily depends on the correlation between the residuals of the two equations \cite{heckman1979sample}. We have taken the logarithm of dependent/independent variables of interest in order to compute elasticity as in \cite{ewing2017does}. For each type of trip modality, we use a maximum likelihood estimation (MLE) framework to fit Eqs.~\ref{model1}-\ref{model2} simultaneously. 

For the trip-level controls, we include controls for individual characteristics (age/gender), whether the household has a motorized vehicle, as well as controls for the purpose of the trip (see Data section). Importantly, Heckman's correction requires that the independent variables in the two equations must not be the exact same. Hence, we drop age, gender, and whether the household has a motorized vehicle in the regression equation Eq.~\ref{model2}.

For the spatial controls, we include variables corresponding to urban form and SES. In practice, we find high multicollinearity between $access(\lambda)$ and density of homes, jobs, and stratum. Accordingly, we reduce the number of variables in order to reduce multicollinearity (see SI for more details and alternative models). Specifically, we include the local cell-level density of homes and jobs\footnote{We use raw density here rather than the previously introduced variables $homes(grav)$ and $jobs(grav)$ as density is commonly used as a trip regression covariate in the literature.} as a composite variable and drop $prox(CBD)$ altogether, noting that combining home and job variables has the effect of proxying for general density in the city; and also keep stratum. Even with this reduced model, we find that multicollinearity can still become problematic (high variance inflation factor) for high $\lambda$\footnote{Essentially, at high values of $\lambda$, $access(\lambda)$ tends towards a smooth distribution capturing the broad urban form rather than localised amenity hotspots.}, hence we limit our analysis to $\lambda\le3$ (see SI for detailed analysis). Additionally, we also report regression results when these spatial controls are excluded, calling this model I and the full model II.

\section{Results}

\subsection{How are amenities distributed?}

Before examining the effect of amenities on urban mobility, we first explore the spatial distribution of the various individual amenity classes throughout the city. Each amenity class obeys a unique spatial distribution throughout the city. Even though all classes broadly depend on the distribution of potential customers throughout the city, we would expect - for example - significant differences between the distribution of convenience stores and physiotherapists. And, as observed in \cite{chong2020economic}, wealthier areas have more diverse amenities (see for instance Fig.~\ref{fig_dists}D-E). 

\begin{figure}
    \centering
    \includegraphics[width=\textwidth]{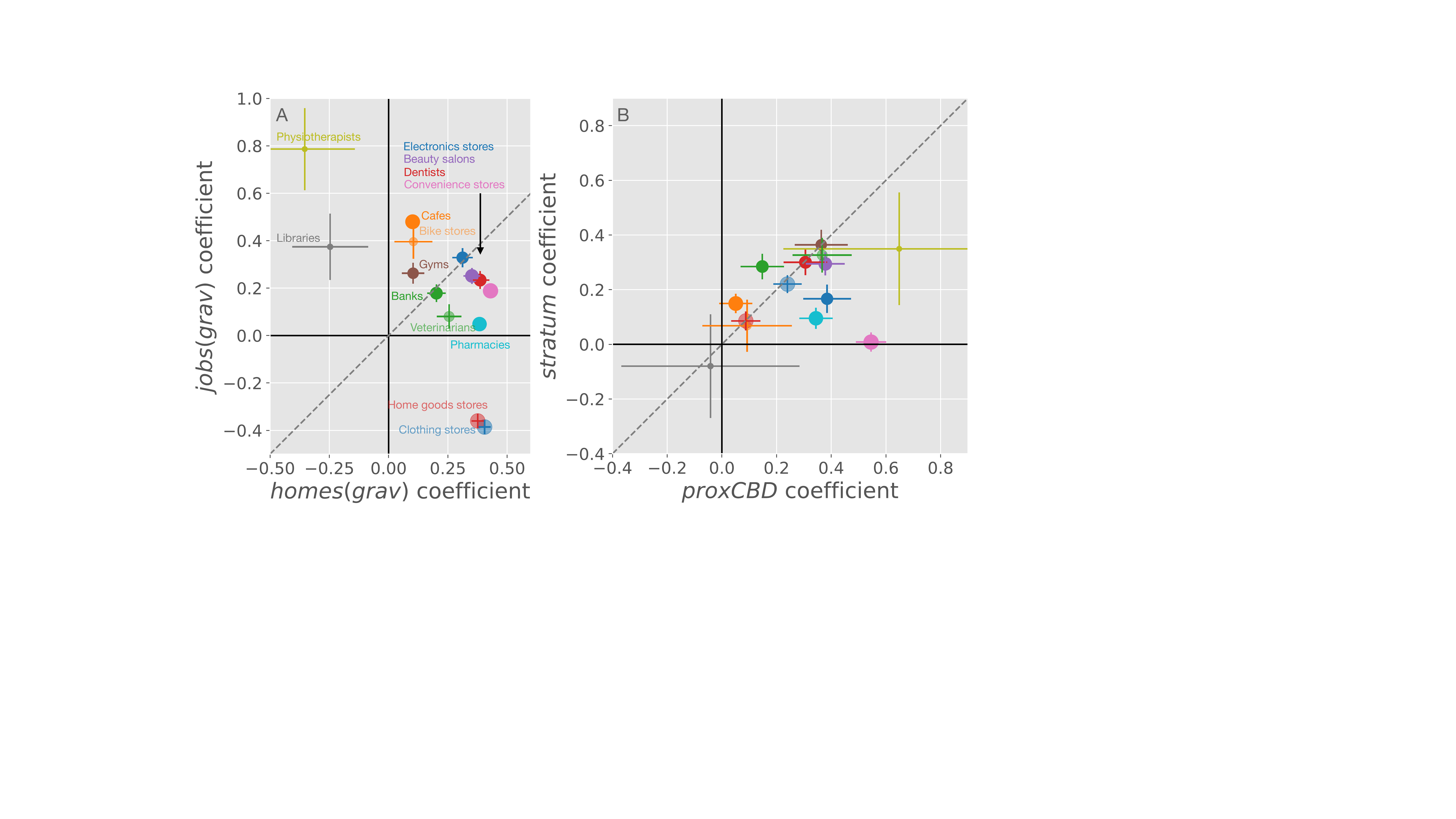}
    \caption{{\bf How do individual amenity classes depend on urban form?} Subfigure A shows the coefficient of homes and jobs intensity with respect to the distribution of amenity classes. For example, we observe clothing stores closely follow home intensity patterns, while physiotherapists co-locate with jobs. Subfigure B shows the coefficient of stratum and proximity to the central business district with respect to the distribution of the various amenity classes.} 
    \label{fig_amenities}
\end{figure}

First, we investigate the correlation between the spatial distribution of homes (and jobs) and each amenity class separately. Our findings illuminate considerable heterogeneity in this relationship. For instance, physiotherapists, libraries, cafes, bike shops, and gyms are more strongly associated with the jobs distribution, while home goods stores, clothing stores, pharmacies, convenience stores, dentists, and beauty salons have a stronger relationship with the homes distribution (Fig.~\ref{fig_amenities}A). Electronics stores and banks have roughly equal association with both homes and jobs. Nearly all amenities have positive associations with SES and proximity to the CBD, with exceptions of libraries, convenience stores, and bicycle stores (Fig.~\ref{fig_amenities}B). Interestingly, gyms, physiotherapists, veterinarians, dentists, and beauty salons have the strongest association with SES, indicating that wealthier residents have greater access to services associated especially with health, well-being, and physical appearance. 

While we observe differences in how the various amenity classes are distributed across the city, our metric $access(\lambda)$ is intended to summarise the composite spatial accessibility to all classes of amenities. However, does the heterogeneity among the amenity distributions render this goal inappropriate? One simple way to answer this question is to simply use standard principal component analysis or PCA to reduce the dimension of the matrix described by the cross-amenity concatenation of $A_z^c$ (interpreting each as a vector) to a vector. Applying this technique, we find that the 1st principal component (PC) explains $61\%$ of the variance across the individual distributions of amenity counts (see SI). Moreover, as we further detail in the SI, while there is variation in the spatial distribution of individual amenity classes, our metric $access(\lambda)$ captures the majority of this variation.

\begin{figure}[t!]
    \includegraphics[width=\textwidth]{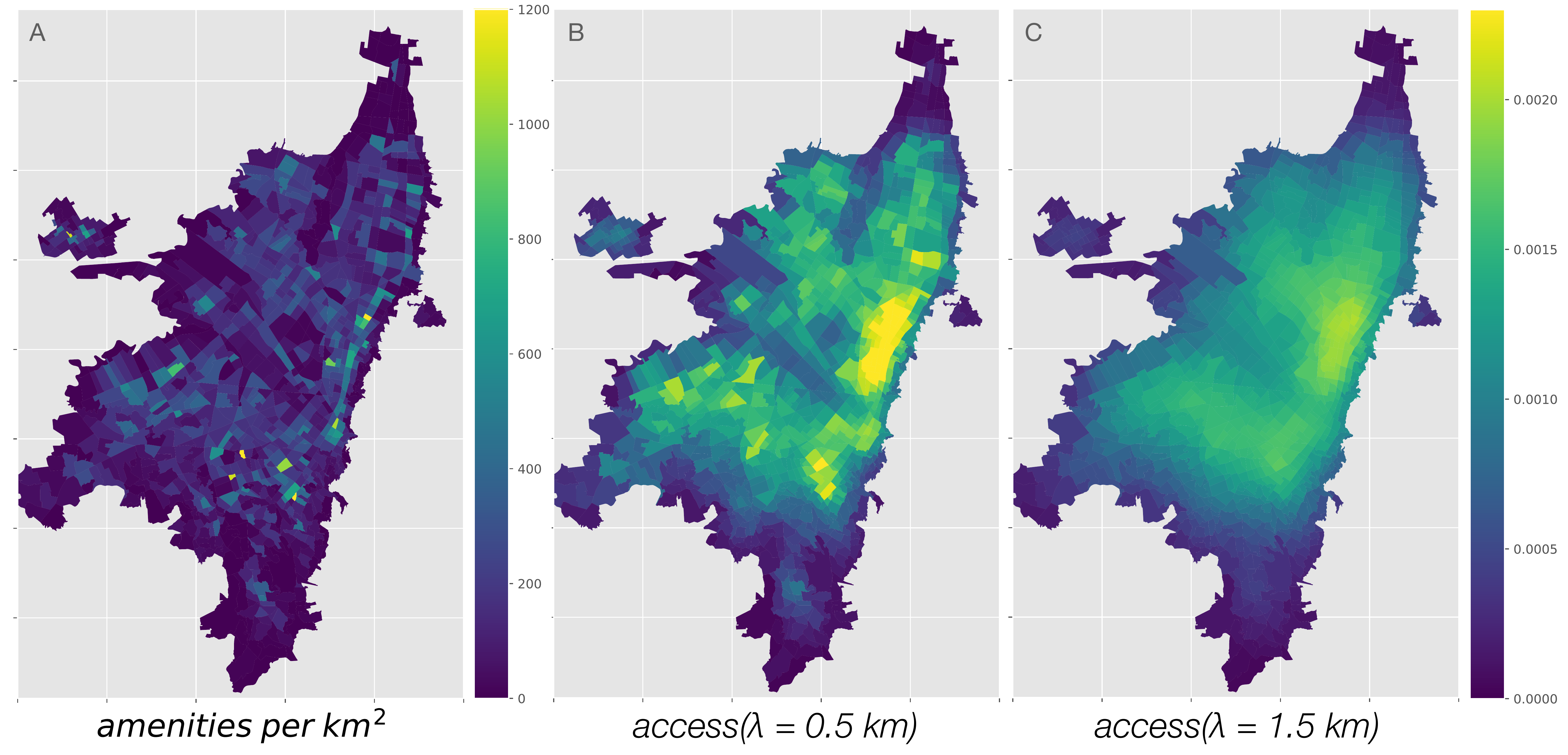}
    \centering
    \caption{{\bf A generalised accessibility metric to capture varied amenity classes and spatial scales.} Subfigure A displays the raw amenity density throughout the city centre, while B-C display the distribution of amenity accessibility at two spatial scales. In B, only nearby amenities contribute substantially to the metric, while more distant contributions give rise to a spatially smoother distribution in C.}
    \label{fig_scales}
\end{figure}

Hence, we can deduce that aggregation across amenity classes can be justified to quantify amenity accessibility in a general sense. We display the spatial distribution of accessibility for varying $\lambda$ in Fig.~\ref{fig_scales}. For low $\lambda$, we observe a very peaky distribution, with very high density all around the city centre but also many local maxima scattered throughout the study area. As we increase $\lambda$, we observe increasing smoothness in the accessibility distribution. Whereas at $\lambda=0.5$ km, we see that amenity hotspots are frequently surrounded by low accessibility zones (e.g., yellow zones surrounded by blue), increasing $\lambda$ to 1.5 km has the effect of spreading out these hotspots into nearby zones.

\subsection{Amenity accessibility impacts mobility behavior}

The primary goal of our work is to further untangle the relationship between accessibility to amenities and urban mobility patterns. For example, does an abundance of local amenities increase our propensity to walk in the local area? Does it reduce the need for long car journeys? We organise our discussion according to trip modes.

\begin{center}
\begin{table}

\tiny\sf\centering
\caption{Effects of $access$ on propensity to walk}
\textbf{Dep. Variable}: $Pr(walk)_t$\\

\begin{tabular}{c | c c| c c| c c| c c }
\toprule

Respondents & {\bf(all)} &  & {\bf(low SES)} & & {\bf(middle SES)} && {\bf(high SES)}&\\ 

\midrule
\textbf{Ind. Variable} & Model I &Model II & Model I&Model II &Model I&Model II & Model I  &Model II\\
 \midrule \\[-1.8ex]
 
$'log\;access_{z(t)}$ &   0.06$^{***}$ &   0.12$^{***}$ &   0.12$^{***}$ &    0.1$^{***}$ &    0.1$^{***}$ &     0.06$^{*}$ &   0.54$^{***}$ &    0.4$^{***}$ \\
                     &         (0.01) &         (0.02) &         (0.02) &         (0.03) &         (0.04) &         (0.04) &         (0.08) &          (0.1) \\
    $stratum_{z(t)}$ &             &  -0.12$^{***}$ &           &           &            &          &         &           \\
                     &          &         (0.01) &          &       &       &       &         &           \\
 $homes+jobs_{z(t)}$ &             &    5.13E-6$^{***}$ &           &            3.01E-6 &            &    5.70E-6$^{***}$ &             &     10.58E-6$^{**}$ \\
                     &           &          (1.26E-6) &           &          (1.98E-6) &           &          (1.83E-6) &          &          (4.63E-6) \\
           Intercept &  -1.17$^{***}$ &  -0.95$^{***}$ &  -1.05$^{***}$ &  -1.16$^{***}$ &  -1.27$^{***}$ &  -1.51$^{***}$ &           0.13 &          -0.43 \\
                     &         (0.06) &          (0.01) &          (0.10) &         (0.14) &          (0.10) &         (0.13) &         (0.26) &         (0.35) \\
         own car$_t$ &  -0.45$^{***}$ &  -0.36$^{***}$ &  -0.27$^{***}$ &  -0.27$^{***}$ &  -0.46$^{***}$ &  -0.45$^{***}$ &  -0.37$^{***}$ &  -0.36$^{***}$ \\
                     &         (0.02) &         (0.02) &         (0.04) &         (0.04) &         (0.03) &         (0.03) &         (0.11) &         (0.11) \\
         weekend$_t$ &  -0.06$^{***}$ &   -0.05$^{**}$ &          -0.05 &          -0.05 &   -0.07$^{**}$ &   -0.07$^{**}$ &          -0.03 &          -0.03 \\
                     &         (0.02) &         (0.02) &         (0.04) &         (0.04) &         (0.03) &         (0.03) &         (0.09) &         (0.09) \\
          female$_t$ &   0.24$^{***}$ &   0.24$^{***}$ &   0.27$^{***}$ &   0.27$^{***}$ &   0.23$^{***}$ &   0.23$^{***}$ &     0.14$^{*}$ &     0.15$^{*}$ \\
                     &         (0.02) &         (0.02) &         (0.04) &         (0.04) &         (0.03) &         (0.03) &         (0.08) &         (0.08) \\
             age$_t$ &    0.0$^{***}$ &    0.0$^{***}$ &    0.0$^{***}$ &    0.0$^{***}$ &   0.01$^{***}$ &   0.01$^{***}$ &            0.0 &            0.0 \\
                     &          (0.0) &          (0.0) &          (0.0) &          (0.0) &          (0.0) &          (0.0) &          (0.0) &          (0.0) \\
            dine$_t$ &   1.53$^{***}$ &   1.59$^{***}$ &   1.84$^{***}$ &   1.84$^{***}$ &   1.63$^{***}$ &   1.64$^{***}$ &   0.73$^{***}$ &   0.74$^{***}$ \\
                     &         (0.06) &         (0.06) &         (0.12) &         (0.12) &         (0.08) &         (0.08) &         (0.17) &         (0.17) \\
          errand$_t$ &   0.78$^{***}$ &   0.79$^{***}$ &   0.71$^{***}$ &   0.71$^{***}$ &   0.86$^{***}$ &   0.87$^{***}$ &   0.69$^{***}$ &   0.69$^{***}$ \\
                     &         (0.04) &         (0.04) &         (0.06) &         (0.06) &         (0.05) &         (0.05) &         (0.14) &         (0.14) \\
      recreation$_t$ &    1.2$^{***}$ &   1.25$^{***}$ &   1.49$^{***}$ &   1.49$^{***}$ &   1.19$^{***}$ &   1.19$^{***}$ &   0.73$^{***}$ &    0.7$^{***}$ \\
                     &         (0.05) &         (0.05) &         (0.09) &         (0.09) &         (0.07) &         (0.07) &         (0.17) &         (0.17) \\
        shopping$_t$ &   1.88$^{***}$ &   1.89$^{***}$ &   2.07$^{***}$ &   2.07$^{***}$ &   1.87$^{***}$ &   1.88$^{***}$ &   1.05$^{***}$ &   1.06$^{***}$ \\
                     &         (0.04) &         (0.04) &         (0.06) &         (0.06) &         (0.05) &         (0.05) &         (0.13) &         (0.13) \\
       religious$_t$ &   1.29$^{***}$ &    1.3$^{***}$ &   1.28$^{***}$ &   1.29$^{***}$ &   1.36$^{***}$ &   1.36$^{***}$ &   0.85$^{***}$ &   0.83$^{***}$ \\
                     &         (0.06) &         (0.06) &         (0.11) &         (0.11) &         (0.09) &         (0.09) &         (0.25) &         (0.26) \\
        physical$_t$ &    1.9$^{***}$ &   1.95$^{***}$ &   1.84$^{***}$ &   1.84$^{***}$ &   2.04$^{***}$ &   2.05$^{***}$ &   1.49$^{***}$ &   1.49$^{***}$ \\
                     &         (0.05) &         (0.05) &         (0.08) &         (0.08) &         (0.07) &         (0.07) &         (0.16) &         (0.16) \\
         search$_t$ &   1.21$^{***}$ &   1.19$^{***}$ &   1.27$^{***}$ &   1.26$^{***}$ &   1.25$^{***}$ &   1.26$^{***}$ &           0.05 &           0.02 \\
                     &         (0.06) &         (0.06) &         (0.08) &         (0.08) &         (0.09) &         (0.09) &          (0.3) &          (0.3) \\\bottomrule \\[-1.8ex] 
                 Obs &        15133 &        15133 &         6005 &         6005 &         7932 &         7932 &         1196 &         1196.0 \\
        Pseudo-$R^2$ &           0.21 &           0.21 &           0.24 &           0.24 &           0.21 &           0.21 &           0.13 &           0.13 \\ \bottomrule

\end{tabular}
\label{pr_walk}
\end{table}
\end{center}
\begin{center}
\begin{table}
\tiny
\caption{Effects of amenity accessibility on walking trip duration}
\textbf{Dep. Variable}: $\log\;duration_{t}$ (walking trips)\\

\begin{tabular}{c | c c |c c| c c| c c }
Respondent subset & (all) & (all) &(low SES)&(low SES)&(middle SES)&(middle SES)&(high SES)&(high SES)\\
\toprule
Model & (I) & (II) & (I) & (II) & (I) & (II) & (I) & (II) \\
\toprule 
  $\log\;access(0.5\;km)_{z(t)}$   &          0.01 &           0.01 &    0.04$^{**}$ &            0.0 &          -0.03 &          -0.03 &  -0.34$^{***}$ &              -0.2 \\
                     &         (0.01) &         (0.02) &         (0.02) &         (0.02) &         (0.03) &         (0.03) &         (0.13) &            (0.14) \\
    $stratum_{z(t)}$ &         &            0.0 &          &          &          &        &         &                \\
                     &          &         (0.01) &          &       &        &         &       &            \\
 $homes+jobs_{z(t)}$ &            &          0.02E-06 &         &        2.1e-06 &            &         -0.7E-08 &             &  -9.87E-06$^{**}$ \\
                     &           &     (1.11E-06) &         &      (1.70E-06) &           &     (1.65E-06) &           &        (4.61E-06) \\
           Intercept &   3.73$^{***}$ &   3.87$^{***}$ &   3.84$^{***}$ &   4.18$^{***}$ &   3.56$^{***}$ &   3.57$^{***}$ &   2.87$^{***}$ &      3.45$^{***}$ \\
                     &         (0.09) &         (0.11) &         (0.13) &         (0.14) &         (0.13) &         (0.16) &         (0.31) &            (0.41) \\
         weekend$_t$ &          -0.02 &          -0.01 &          -0.01 &          -0.01 &          -0.03 &          -0.03 &           0.03 &              0.04 \\
                     &         (0.02) &         (0.02) &         (0.03) &         (0.03) &         (0.03) &         (0.03) &         (0.08) &            (0.08) \\
            dine$_t$ &  -1.02$^{***}$ &  -1.13$^{***}$ &  -0.99$^{***}$ &  -1.35$^{***}$ &   -1.0$^{***}$ &  -1.01$^{***}$ &   -0.7$^{***}$ &      -0.7$^{***}$ \\
                     &         (0.08) &         (0.08) &         (0.14) &         (0.12) &         (0.11) &         (0.11) &         (0.24) &            (0.23) \\
          errand$_t$ &  -0.45$^{***}$ &   -0.5$^{***}$ &  -0.42$^{***}$ &  -0.59$^{***}$ &  -0.43$^{***}$ &  -0.43$^{***}$ &   -0.42$^{**}$ &      -0.42$^{**}$ \\
                     &         (0.06) &         (0.06) &         (0.09) &         (0.08) &         (0.08) &         (0.08) &         (0.21) &             (0.2) \\
      recreation$_t$ &  -0.74$^{***}$ &  -0.82$^{***}$ &  -0.91$^{***}$ &  -1.22$^{***}$ &  -0.53$^{***}$ &  -0.54$^{***}$ &  -0.78$^{***}$ &     -0.73$^{***}$ \\
                     &         (0.07) &         (0.07) &         (0.12) &         (0.11) &          (0.1) &          (0.1) &         (0.24) &            (0.24) \\
        shopping$_t$ &  -1.14$^{***}$ &  -1.25$^{***}$ &  -1.19$^{***}$ &   -1.6$^{***}$ &  -1.04$^{***}$ &  -1.05$^{***}$ &  -0.74$^{***}$ &     -0.75$^{***}$ \\
                     &         (0.07) &         (0.07) &         (0.12) &         (0.09) &          (0.1) &          (0.1) &         (0.26) &            (0.25) \\
       religious$_t$ &  -0.78$^{***}$ &  -0.87$^{***}$ &  -0.71$^{***}$ &  -1.01$^{***}$ &  -0.75$^{***}$ &  -0.75$^{***}$ &  -0.84$^{***}$ &     -0.81$^{***}$ \\
                     &         (0.08) &         (0.08) &         (0.13) &         (0.12) &         (0.11) &         (0.11) &         (0.31) &            (0.31) \\
        physical$_t$ &  -0.91$^{***}$ &  -1.03$^{***}$ &  -0.87$^{***}$ &  -1.23$^{***}$ &  -0.86$^{***}$ &  -0.87$^{***}$ &   -0.74$^{**}$ &     -0.75$^{***}$ \\
                     &         (0.07) &         (0.07) &         (0.12) &          (0.1) &         (0.11) &         (0.11) &         (0.31) &            (0.29) \\
         search$_t$ &  -0.85$^{***}$ &  -0.93$^{***}$ &  -0.88$^{***}$ &  -1.16$^{***}$ &  -0.78$^{***}$ &  -0.78$^{***}$ &    -0.65$^{*}$ &             -0.58 \\
                     &         (0.08) &         (0.07) &         (0.11) &          (0.1) &         (0.11) &         (0.11) &         (0.38) &            (0.38) \\
           $\Lambda$ &          -0.29$^{***}$ &          -0.38$^{***}$ &          -0.26$^{***}$ &          -0.57$^{***}$ &          -0.24$^{***}$ &          -0.24$^{***}$ &          -0.39 &             -0.39 \\
                     &         (0.05) &         (0.05) &         (0.08) &         (0.07) &         (0.06) &         (0.06) &         (0.35) &            (0.35) \\\bottomrule \\[-1.8ex] 
          Obs(total) &        15133 &        15133 &         6005 &         6005 &         7932 &         7932 &         1196 &            1196 \\
       Obs(censored) &       7257 &       7257) &       3161 &       3161 &       3643 &       3643 &        453&           453 \\
               $R^2$ &           0.07 &           0.07 &           0.09 &           0.09 &           0.06 &           0.06 &           0.05 &              0.06 \\ \bottomrule

\end{tabular}
\label{trips_walk}
\end{table}
\end{center}

\subsubsection{Walking}

First, we fit Eqn \ref{model1} for the probabilities of using various trip modes. Initially, we select a single value of $\lambda$, our spatial scale parameter, choosing $\lambda=0.5$ km. We find that higher amenity accessibility is associated with a higher probability of walking for the full sample (columns 1-2 of Table \ref{pr_walk}), with even stronger results in model II wherein SES and residential/work density are included as controls. Considering our control variables, we find a negative effect for stratum, indicating that lower SES people are in general more likely to walk (as in \cite{guzman2017urban}). Unsurprisingly, we find a negative effect for car ownership. Density of homes and jobs is on the other hand associated with increased walking overall, as in other studies \cite{ewing2001travel}. Women are likelier to walk, perhaps due to childcare and household load \cite{duchene2011gender}. 

\begin{figure}
    \centering
    \includegraphics[width=\textwidth]{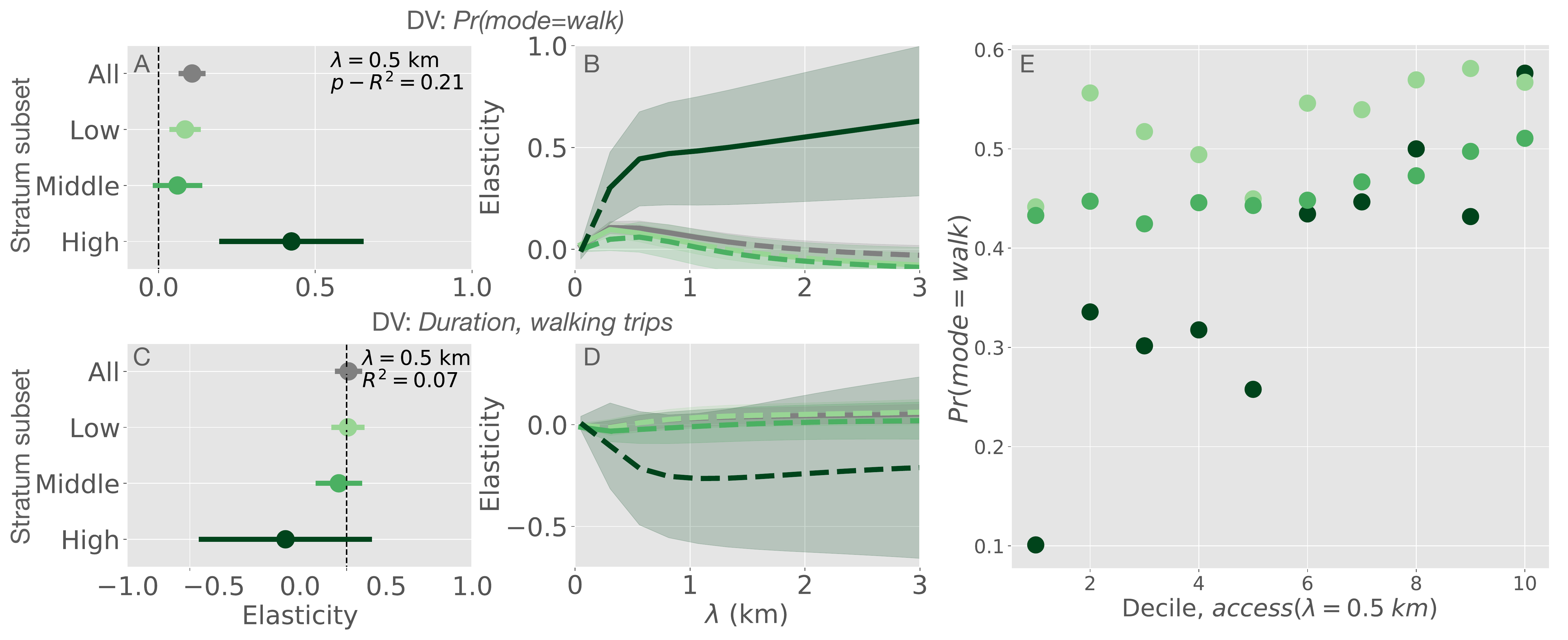}
    \caption{{\bf Effects of accessibility on tendency to make walking trips (first row) and walking trip druation (bottom).} The first column displays the elasticities of $\log \;access(\lambda)$ at $\lambda=0.5\;km$ for models I and II (without/with spatial controls) within both the full sample of respondents and then also for SES subgroups. In the second column, we display coefficients/confidence intervals for the full sample as well as SES subgroups (grey corresponding to all respondents, light green corresponding to elasticity within the low SES subset, etc.). Solid lines indicate significance ($p<.05$), whereas dashed lines indicate insignificance. In the right panel, we plot the probability of walking across increasing deciles of accessibility for each SES group.}
    \label{figreg_w}
\end{figure}

Next, we wish to better understand how amenities impact the tendency to make walking trips for specific SES groups. For example, does having amenities nearby have a different influence on trip behaviour for poorer vs richer citizens? We divide our observations (trips) into three groups by the SES of their home location: stratum 1/2 (low SES), stratum 3/4 (middle), and stratum 5/6 (high). We then drop the control for stratum and investigate model I and II for the set of trips originating in each group. We both present the results in columns 3-8 of Table \ref{pr_walk} as well as depict the results visually for model II in the upper left panel of Figure \ref{figreg_w} (corresponding figures for model I are found in the SI).

We find that effects of amenities on the likelihood of walking are strongest for the wealthiest group. At $\lambda=0.5$ km (Model II), a $1\%$ increase in accessibility results in a $0.40\%$ increase in the likelihood of walking vs about a $0.06-0.12\%$ increase for the lower and middle SES groups (left column of Fig. \ref{figreg_w}). Moreover, the coefficient for the middle SES group is only significant at the $p<.1$ level. Results are fairly similar in model I, with the only major difference being that the coefficients for all three groups are significant in this model.

Our results point toward an important role for amenities in the mobility behaviour of high SES residents, but less so for low/middle SES residents. In order to illustrate this, we compute $Pr(walk)$ for each decile of accessibility for each SES group in the right panel of Fig.~\ref{figreg_w}. For the high SES group, we find that the probability to walk clearly increases with accessibility. In contrast, we observe at best a weak pattern for the low and middle SES groups. 

We also consider variation in the spatial scale parameter $\lambda$ for the effect of amenity accessibility on walking trip propensity. Specifically, we examine the elasticity of amenity accessibility as a function of $\lambda$ on trip mode propensity (see second column of Fig.~\ref{figreg_w}). A peak for smaller values of $\lambda$ implies that only closeby amenities are associated with trip behavior, whereas a peak for a larger value means that amenity hotspots within the general area of the city play a role. 

The high SES group exhibits a strong association between amenity accessibility and the probability of walking. The coefficient for this group increases significantly between $\lambda=0.1-0.7$ km, and then more or less levels off with no further significant increase. This suggests that nearby amenities heavily influence walking for this group, while more distant amenities have little additional influence. The coefficient for the other two groups peak for small $\lambda,$ and both become insignificant for higher $\lambda.$ Again, this suggests that only nearby amenities influence the propensity to walk for these groups.

In addition, we investigate effects on walking duration - which we fit according to Eq.~\ref{model2} simultaneously with Eq.~\ref{model1} via MLE. Again, we first choose $\lambda=0.5$ km and later investigate effects for varying $\lambda$, reporting results in Table~\ref{trips_walk} and depicting them visually in Fig.~\ref{figreg_w}. The coefficient of accessibility for the full sample is insignificant for all $\lambda$ - suggesting that while amenities do influence people to walk, they do not reduce the length of the walking trip in general. This is not so surprising: walkers in amenity-rich areas may not seek to minimise time as they are more likely to be taking recreational trips and visiting multiple amenities (trip chaining). Nor do we find significant effects for stratum or for density of homes and jobs. We do find significant effects for the inverse mills ratio, indicating that modeling walking propensity and walking duration independently (i.e., without our use of Heckman's correction) would result in selection bias.

We do find some differences in the effects of accessibility on walking duration for the various socioeconomic groups. That is, we find that there is no significant effect within the low/middle SES groups, but find at least mixed evidence for an effect in the high SES group. Specifically, we find significant effects for the high SES group at the $p<.05$ level in model I, but the effect is only significant at the $p<.1$ level when we control for residential and work density. These findings seem to hold across our range of $\lambda$. Hence, higher amenity accessibility is loosely associated with shorter walking trips for high SES residents, but has no effect on walking duration for the low and middle SES groups. 

Overall, we find that higher amenity accessibility is associated with a higher probability of walking but has very little if any effect on walking trip duration. By far, the strongest effects for walking propensity emerge for the wealthiest group. The coefficient for the high SES group tends to increase until about 0.7 km, at which point it levels off. 

\begin{center}
\begin{table}
\tiny
\caption{Effects of $access$ on bus trip duration}
\textbf{Dep. Variable}: $\log\;duration_{t}$ (bus trips)\\

\begin{tabular}{c | c c| c c |c c| c c }
Respondent subset & (all) & (all) &(low SES)&(low SES)&(middle SES)&(middle SES)&(high SES)&(high SES)\\

\toprule
Model & (I) & (II) & (I) & (II) & (I) & (II) & (I) & (II) \\
\toprule \\
 $\log\;access(1.5\;km)_{z(t)}$  &    -0.15$^{***}$ &     -0.15$^{***}$ &  -0.11$^{***}$ &     -0.17$^{***}$ &  -0.26$^{***}$ &  -0.26$^{***}$ &          0.05 &          0.15 \\
                     &         (0.02) &            (0.02) &         (0.02) &            (0.03) &         (0.05) &         (0.05) &        (0.09) &        (0.13) \\
    $stratum_{z(t)}$ &             &     -0.05$^{***}$ &          &            &           &          &           &          \\
                     &          (0.0) &            (0.01) &          (0.0) &             (0.0) &          (0.0) &          (0.0) &         (0.0) &         (0.0) \\
 $homes+jobs_{z(t)}$ &             &  3.59E-06$^{***}$ &          &  5.28e-06$^{***}$ &       &        1.4E-06 &         &     -8.28E-06 \\
                     &          &        (1.18E-06) &         &        (1.69E-06) &         &     (1.74E-06) &         &    (6.55E-06) \\
           Intercept &   3.88$^{***}$ &      3.78$^{***}$ &   4.08$^{***}$ &      3.68$^{***}$ &   3.67$^{***}$ &   3.61$^{***}$ &  4.12$^{***}$ &  4.57$^{***}$ \\
                     &         (0.09) &            (0.11) &         (0.11) &            (0.13) &         (0.17) &          (0.2) &        (0.41) &        (0.57) \\
         weekend$_t$ &           0.01 &              0.02 &    0.08$^{**}$ &       0.08$^{**}$ &    -0.06$^{*}$ &          -0.05 &         -0.01 &         -0.03 \\
                     &         (0.02) &            (0.02) &         (0.03) &            (0.03) &         (0.03) &         (0.03) &        (0.13) &        (0.13) \\
            dine$_t$ &  -0.41$^{***}$ &     -0.56$^{***}$ &   -0.8$^{***}$ &     -0.94$^{***}$ &          -0.13 &          -0.15 &         -0.09 &         -0.03 \\
                     &         (0.14) &            (0.14) &         (0.24) &            (0.23) &          (0.2) &          (0.2) &         (0.4) &         (0.4) \\
          errand$_t$ &  -0.13$^{***}$ &     -0.16$^{***}$ &  -0.21$^{***}$ &     -0.24$^{***}$ &          -0.05 &          -0.06 &          0.17 &          0.13 \\
                     &         (0.03) &            (0.03) &         (0.04) &            (0.04) &         (0.04) &         (0.04) &        (0.15) &        (0.15) \\
      recreation$_t$ &  -0.22$^{***}$ &     -0.28$^{***}$ &    -0.19$^{*}$ &      -0.27$^{**}$ &    -0.16$^{*}$ &    -0.18$^{*}$ &          0.19 &          0.18 \\
                     &         (0.07) &            (0.07) &         (0.12) &            (0.11) &          (0.1) &          (0.1) &        (0.19) &        (0.19) \\
        shopping$_t$ &  -0.37$^{***}$ &     -0.51$^{***}$ &  -0.55$^{***}$ &     -0.68$^{***}$ &          -0.16 &          -0.17 &           0.1 &          0.13 \\
                     &         (0.08) &            (0.08) &         (0.12) &            (0.11) &         (0.12) &         (0.13) &        (0.21) &        (0.21) \\
       religious$_t$ &   -0.16$^{**}$ &     -0.23$^{***}$ &  -0.37$^{***}$ &     -0.45$^{***}$ &           -0.0 &          -0.01 &          0.47 &          0.47 \\
                     &         (0.08) &            (0.08) &         (0.12) &            (0.12) &          (0.1) &         (0.11) &        (0.32) &        (0.32) \\
        physical$_t$ &  -0.47$^{***}$ &     -0.61$^{***}$ &  -0.65$^{***}$ &      -0.8$^{***}$ &    -0.26$^{*}$ &    -0.28$^{*}$ &          0.06 &          0.17 \\
                     &          (0.1) &             (0.1) &         (0.15) &            (0.14) &         (0.16) &         (0.16) &        (0.43) &        (0.44) \\
         search$_t$ &   -0.2$^{***}$ &     -0.28$^{***}$ &   -0.22$^{**}$ &     -0.31$^{***}$ &          -0.16 &          -0.17 &         -0.18 &         -0.19 \\
                     &         (0.07) &            (0.07) &          (0.1) &            (0.09) &         (0.11) &         (0.12) &        (0.32) &        (0.32) \\
           $\Lambda$ &           0.05 &               0.20 &           0.04 &              0.17 &           -0.00 &           0.02 &          0.06 &          0.06 \\
                     &         (0.07) &            (0.08) &         (0.09) &            (0.09) &         (0.13) &         (0.13) &        (0.19) &        (0.19) \\\bottomrule \\[-1.8ex] 
          Obs(total) &        15133 &           15133 &         6005 &            6005 &         7932 &         7932 &        1196 &        1196 \\
       Obs(censored) &       3689 &          3689 &       1771 &          1771 &       1804 &       1804 &       114 &       114 \\
               $R^2$ &           0.05 &              0.05 &           0.07 &              0.08 &           0.03 &           0.03 &          0.04 &          0.06\\ \bottomrule \end{tabular}
\label{trips_bus}
\end{table}
\end{center}

\begin{center}
\begin{table}
\tiny
\caption{Effects of $access$ on car/moto trip duration}
\textbf{Dep. Variable}: $\log\;duration_{t}$ (car/moto trips)\\

\begin{tabular}{c | c c| c c| c c |c c }
Respondent subset & (all) & (all) &(low SES)&(low SES)&(middle SES)&(middle SES)&(high SES)&(high SES)\\
\toprule
Optimal $\lambda$\\
\toprule
Model & (I) & (II) & (I) & (II) & (I) & (II) & (I) & (II) \\
\toprule \\
$\log\;access(1.5\;km)_{z(t)}$ & -0.12$^{***}$ &    -0.15$^{***}$ &          -0.01 &           -0.0 &    -0.14$^{*}$ &   -0.15$^{**}$ &  -0.21$^{***}$ &  -0.28$^{***}$ \\
                     &         (0.04) &           (0.04) &         (0.07) &         (0.08) &         (0.07) &         (0.07) &         (0.07) &         (0.08) \\
    $stratum_{z(t)}$ &           &            -0.01 &          &          &     &   &       &      \\
                     &         &           (0.02) &        &         &         &          &          &      \\
 $homes+jobs_{z(t)}$ &             &  4.69E-06$^{**}$ &          &         -1.00E-06 &           &       3.72E-06 &             &       7.05E-06 \\
                     &           &       (1.99E-06) &       &     (4.45E-06) &         &     (2.66E-06) &   &     (4.82E-06) \\
           Intercept &   3.44$^{***}$ &     3.22$^{***}$ &   3.99$^{***}$ &   4.05$^{***}$ &   3.28$^{***}$ &   3.17$^{***}$ &   2.79$^{***}$ &   2.47$^{***}$ \\
                     &         (0.15) &           (0.17) &         (0.24) &         (0.33) &         (0.18) &         (0.19) &         (0.29) &         (0.26) \\
         weekend$_t$ &     0.06$^{*}$ &             0.06 &           0.02 &           0.02 &     0.08$^{*}$ &     0.08$^{*}$ &           0.07 &           0.05 \\
                     &         (0.04) &           (0.04) &         (0.08) &         (0.08) &         (0.05) &         (0.05) &         (0.09) &         (0.09) \\
            dine$_t$ &  -0.56$^{***}$ &    -0.56$^{***}$ &          -0.32 &          -0.33 &  -0.67$^{***}$ &  -0.67$^{***}$ &  -0.38$^{***}$ &  -0.39$^{***}$ \\
                     &         (0.08) &           (0.08) &         (0.22) &         (0.22) &          (0.1) &          (0.1) &         (0.14) &         (0.14) \\
          errand$_t$ &  -0.24$^{***}$ &    -0.24$^{***}$ &  -0.33$^{***}$ &  -0.33$^{***}$ &  -0.27$^{***}$ &  -0.27$^{***}$ &           -0.1 &          -0.11 \\
                     &         (0.05) &           (0.05) &         (0.11) &         (0.11) &         (0.06) &         (0.06) &         (0.12) &         (0.12) \\
      recreation$_t$ &  -0.29$^{***}$ &    -0.28$^{***}$ &  -0.47$^{***}$ &  -0.48$^{***}$ &  -0.39$^{***}$ &  -0.39$^{***}$ &           0.05 &           0.03 \\
                     &         (0.07) &           (0.07) &         (0.18) &         (0.18) &          (0.1) &          (0.1) &         (0.16) &         (0.16) \\
        shopping$_t$ &  -0.69$^{***}$ &    -0.68$^{***}$ &  -0.59$^{***}$ &   -0.6$^{***}$ &  -0.78$^{***}$ &  -0.77$^{***}$ &  -0.52$^{***}$ &  -0.52$^{***}$ \\
                     &         (0.05) &           (0.05) &         (0.13) &         (0.13) &         (0.07) &         (0.07) &         (0.12) &         (0.12) \\
       religious$_t$ &  -0.36$^{***}$ &    -0.36$^{***}$ &   -0.51$^{**}$ &   -0.51$^{**}$ &   -0.29$^{**}$ &   -0.29$^{**}$ &   -0.7$^{***}$ &   -0.65$^{**}$ \\
                     &         (0.11) &           (0.11) &         (0.23) &         (0.23) &         (0.13) &         (0.13) &         (0.27) &         (0.27) \\
        physical$_t$ &   -0.7$^{***}$ &     -0.7$^{***}$ &  -0.94$^{***}$ &  -0.94$^{***}$ &  -0.67$^{***}$ &  -0.66$^{***}$ &  -0.73$^{***}$ &  -0.73$^{***}$ \\
                     &         (0.09) &           (0.09) &         (0.26) &         (0.26) &         (0.13) &         (0.13) &         (0.18) &         (0.18) \\
         search$_t$ &  -0.49$^{***}$ &    -0.49$^{***}$ &  -0.55$^{***}$ &  -0.55$^{***}$ &  -0.52$^{***}$ &  -0.52$^{***}$ &          -0.29 &           -0.3 \\
                     &         (0.08) &           (0.08) &         (0.16) &         (0.16) &         (0.11) &         (0.11) &         (0.23) &         (0.23) \\
           $\Lambda$ &           0.12** &             0.12** &          -0.02 &          -0.02 &           0.19* &           0.18** &           0.37 &           0.37* \\
                     &         (0.06) &           (0.06) &         (0.09) &         (0.08) &         (0.08) &         (0.08) &         (0.25) &         (0.22) \\\bottomrule \\[-1.8ex]
          Obs(total) &        15133 &              15133 &         6005 &         6005 &         7932 &         7932 &         1196 &         1196 \\
       Obs(censored) &       2245 &         2245 &        427 &        427 &       1364 &       1364 &        454 &        (454.0)  \\
               $R^2$ &            0.1 &                  0.1 &           0.08 &           0.08 &           0.11 &           0.11 &           0.11 &           0.13  \\ \bottomrule
\multicolumn{8}{r}{$^{*}$p$<$0.1; $^{**}$p$<$0.05; $^{***}$p$<$0.01} 
\end{tabular}
\label{trips_car}
\end{table}
\end{center}

\subsubsection{Car \& motorcycle}

How does the presence of amenities influence the probability of taking other modes of transport besides walking? If amenities increase the probability of walking, then they decrease the probability of not walking, e.g., travelling by car or bus. In the SI, we show that this is the case: amenity accessibility is (for all spatial scales) associated with lower probability of taking the car (or bus). As above, this effect is particularly strong for the high SES group. 

\begin{figure}
    \centering
    \includegraphics[width=.67\textwidth]{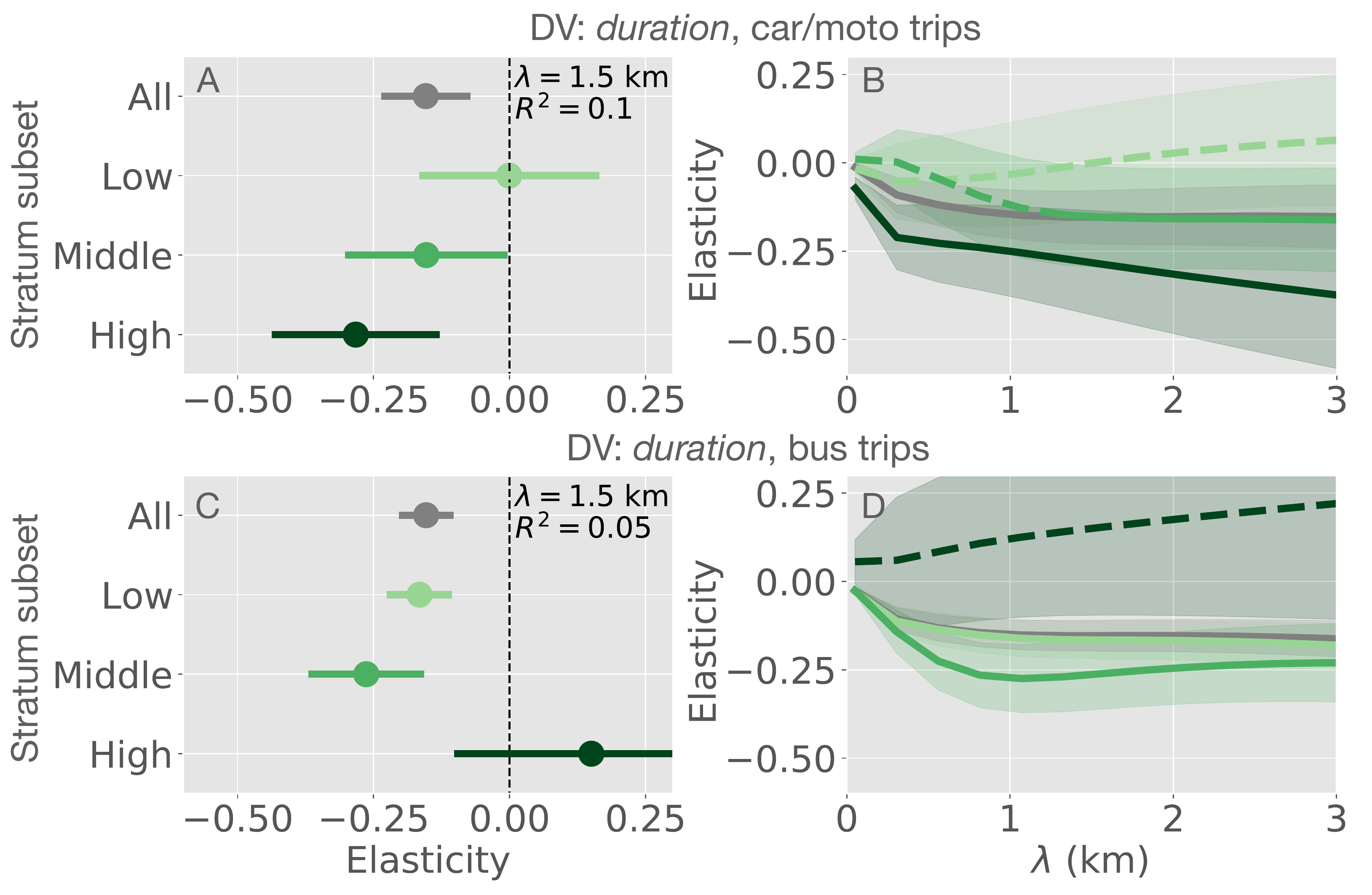}
    \caption{{\bf Effects of amenity accessibility on bus and auto trip duration.} 
    Rows correspond to effects on the various dependent variables. For each row, the left column displays the elasticity/confidence intervals for $\log\;access$ effects at fixed $\lambda$ both for all respondents and then for low/middle/high SES groups. The right column displays the elasticity when we let $\lambda$ vary across a range of spatial scales.}
    \label{figreg}
\end{figure}

We focus here on car trip duration, displaying results in Table~\ref{trips_car} and in the bottom row of Fig.~\ref{figreg}. As above, we first select a single value of our spatial scale parameter $\lambda$. We choose $\lambda=1.5$ km as we a priori expect a larger spatial scale to be more appropriate for investigating bus trip behaviour. At $\lambda=1.5$ km, we find that in the full sample accessibility to amenities is associated with reduced car trip duration. The elasticity is fairly similar whether or not we control for stratum and density of jobs and homes ($-0.12/-0.15$ for models I/II). Higher residential and job density is associated with longer car trip duration, likely signalling effects of congestion. However, we observe no significant effect of SES. Here, we do observe a significant effect for the inverse mills ratio.

Disaggregating effects by SES group, we find that at $\lambda=1.5$ km, amenity accessibility is only associated at the $p<.05$ level with reduced car trips for the upper and middle SES groups. The effect is substantially stronger for the high SES group. When we vary $\lambda$, we see a robust effect for the high SES group but the middle SES group coefficient is only significant for $\lambda>1.2$. As above, both of these coefficients initially increase in magnitude for increasing $\lambda$, but gradually level off. The low SES group coefficient is not significant for any $\lambda$. 

Hence, it seems that the association of amenity accessibility and car trip duration is modulated by SES. That is, for high and perhaps middle SES residents, having more amenities nearby is associated with reduced car trip duration. However, there is no significant association between amenities and car trip duration for the low SES group.

\subsubsection{Bus}

Regarding bus trips, we find that higher amenity accessibility is associated with shorter bus trips, with or without controls for stratum and residential/job density (see Table~\ref{trips_bus} and the first row of Fig.~\ref{figreg}). In fact, our results are strikingly similar between model I and II here, with the elasticity of amenity accessibility being $-0.15$ in both cases. Hence, bus trips beginning in areas of the city with higher amenity accessibility tend to be shorter. SES has a robust negative effect on bus trip duration, meaning that higher stratum residents take shorter bus trips. On the other hand, the density of jobs and homes is associated with increased trip length, perhaps on account of denser areas having more traffic congestion. Effects for trip purpose are generally significant, but we do not find a significant effect of the inverse mills ratio ($\Lambda$), so there seems to be little selection bias in the case of bus trips.

When we examine specific SES groups, we see that this negative effect on bus trip duration holds within the low and middle SES groups, but does not extend to the high SES group. Next, we vary the spatial scale parameter $\lambda$. Focusing first on the full sample (grey), the coefficient decreases steadily with increasing $\lambda$ until about 1 km, at which point the coefficient seems to level off. This suggests that amenities within about 1 km are most associated with lower bus trip duration. Looking at the individual SES groups, we see that for low/middle SES groups, the coefficient tends to exhibit similar behaviour as the full sample model, while the upper SES coefficient is insignificant for all $\lambda$. The middle SES group coefficient tends overall to be slightly stronger than the lower SES group coefficient.

Here, we find that amenities are associated with reduced bus trip duration for the low and middle SES groups, but not for the high SES group. As is the case with car trip duration, effects on bus trip duration tend to increase in magnitude with the spatial scale $\lambda$ until around 1 km before leveling off.


\section{Discussion}

In this paper, we investigate the role of socio-economic status in the relationship between local amenities and mobility patterns. In particular, we leverage data from a large internet repository of location data on amenities, and use this to define an amenity accessibility measure. Our results show that on the whole, having more amenities nearby leads to a modest reduction in car and bus travel distance and increases residents' tendency to walk, in line with other recent work on amenities in Sweden \cite{haugen2013divergent, ellder2020kind, ellder2020local}. However, SES seems to modulate this relationship in an important and hitherto unrecognized manner. Specifically, we find that the presence of amenities increases walking trip propensity and decreases car travel duration for the wealthiest group only. Effects are either insignificant or very small in magnitude for the lower and middle SES groups. On the other hand, amenities tend to reduce bus trip duration (which has been studied less in this literature) for just the low and middle income SES groups.

We suspect that different socioeconomic groups make travel mode decisions based on different factors, particularly in the case of walking. In general, low income people are thought to choose walking for different reasons than wealthy people \cite{boarnet2001travel, mondschein2018persistent}. For example, lower income people may see the cost of driving or taking a bus as a more costly alternative, and factor this cost more heavily into their decision to walk. A longer walking trip may be more preferable than paying a bus fare or paying for gas. By contrast, this cost weighs less heavily on the wealthy, and their decision to walk may depend less on cost and more on proximity. Wealthier suburban residents - with relatively few amenities nearby - can much more efficiently get what they need by driving than by walking. By contrast, inner-city high SES residents live in amenity-rich ``walkable'' neighbourhoods and are hence much likelier to walk. This narrative is consistent with our results which highlight the much stronger effect of amenities on the tendency of the wealthy to walk relative to the low and middle SES groups.

Turning to car trips, we also found that wealthier residents reduced the duration of their trips in response to local amenities. We suspect this result may have a related explanation to that of walking. That is, high SES motorists may seek primarily to minimise time in traffic, whereas lower SES motorists may have have other motivations. In particular, we suspect that this result may be related to a mismatch between amenities in lower SES areas and the needs of the people living there. Our measure only captures average accessibility across amenities, but low income people may be more separated from the specific amenities that they need or wish to access. For instance, previous research has shown that poorer neighbourhoods have lower accessibility to banking and more access to alternative financial institutions (e.g. payday lenders) in the US \cite{small2021banks}, so residents needing, for example, mortgage advice would need to travel further. Additionally, it is well known that poorer areas are often associated with more fast food and fewer supermarkets \cite{sharkey2009association}, and so local residents in amenity-rich areas may be travelling further for their own needs. 

Overall, our results suggest that poorer people's walking and driving behaviour seems to depend less on their local environment relative to wealthy residents. This result somewhat contrasts previous characterisations of the poor as being more connected to their local environment \cite{marquet2017local}. For instance, in the context of urban health, many studies show that the relationship between environment and health is stronger for poorer people \cite{vallee2021everyday}.

Our measure can be adapted to measure proximity to amenities at various spatial scales. A priori we might expect that closeby amenities would influence the propensity to walk whereas further away amenities might impact car trip duration. However, in our analysis we find similar behaviour with regards to spatial scale for all three modes studied. Specifically, the coefficients peak in a short range of 0.5-1km, and then typically level off. We could interpret this as suggesting, for example in the case of car trip duration, is that what matters is not so much the presence of further away amenities, but the absence of nearby amenities. This topic merits further study and perhaps more sophisticated framework for amenity accessibility that takes into account local vs regional amenities as well (similar to \cite{haugen2013divergent}).

Overall, our findings generate important nuance in the debate on the development of more walkable neighbourhoods via bringing residents closer to amenities ("15 minute cities'' or "walkable mixed use communities''). While our study is cross-sectional\footnote{Many studies suggest residential self-selection may limit the inference of cross-sectional studies on transport behaviour - however, recent work suggests it to be relatively minor in importance \cite{naess2014tempest}}, and limited to just one aspect of walkability, our results suggest that adding amenities to a low income vs. a wealthy neighbourhood may have different effects on residents' mobility behaviour. In a wealthy area, amenities alone may drive residents to walk much more, but other neighbourhood improvements may be more impactful for lower and middle income communities. An important avenue for future work is to further investigate the factors that drive lower income residents' mobility behaviour, but quantitatively this is a challenging task at the urban scale as measures that quantify variables such as public safety and street quality are often heavily correlated with SES. Another avenue for future work might be to investigate the impact of factors such as public safety and streetscape quality which can also affect trip behaviour, and differ by SES group. 

\bibliography{references}

\end{document}